\documentclass[5p,times]{elsarticle}

\usepackage[numbers]{natbib}
\usepackage{orcidlink}
\usepackage{float}
\usepackage{subfig}
\usepackage{graphicx}
\usepackage{url}
\usepackage{xcolor}
\usepackage{stfloats}
\usepackage{tcolorbox}
\usepackage{amsmath}
\usepackage{amsfonts}
\usepackage{multirow}
\usepackage{booktabs}
\usepackage{listings}
\usepackage{pifont}
\usepackage{makecell}
\usepackage{nicematrix}
\usepackage{bm}
\usepackage{arydshln} 
\usepackage{threeparttable}
\usepackage{enumitem}

\definecolor{verylightgray}{rgb}{.97,.97,.97}

\usepackage{algorithm}
\usepackage{algpseudocode}

\journal{Information and Software Technology}

\begin{document}

\begin{frontmatter}
	
	\title{Improving vulnerability type prediction and line-level detection via adversarial training-based data augmentation and multi-task learning}

        \author[NTU]{Siyu Chen\orcidlink{0009-0006-8951-4922}}
	\ead{chensiyu043@gmail.com}

        \author[NTU]{Jiongyi Yang\orcidlink{0009-0004-2585-7147}}
        \ead{yjy2004@stmail.ntu.edu.cn}
        
	\author[NTU]{Xiang Chen\orcidlink{0000-0002-1180-3891}\corref{mycorrespondingauthor}}
        \cortext[mycorrespondingauthor]{Corresponding author}
	\ead{xchencs@ntu.edu.cn}	
         
         \author[NTU]{Menglin Zheng\orcidlink{0009-0006-3197-8863}}
	\ead{2230110478@stmail.ntu.edu.cn}
    
         \author[NTU]{Minnan Wei\orcidlink{0009-0007-5479-5784}}
	\ead{minnanvvei@gmail.com}

        \author[NTU]{Xiaolin Ju\orcidlink{0000-0003-2579-5359}}
        \ead{ju.xl@ntu.edu.cn}
    		
	\address[NTU]{School of Artificial Intelligence and Computer Science, Nantong University, Nantong, China}

\begin{abstract}

\textbf{Context:} Software vulnerabilities pose a significant threat to modern software systems, as evidenced by the growing number of reported vulnerabilities and cyberattacks. These escalating trends underscore the urgent need for effective approaches that can automatically detect and understand software vulnerabilities.

\textbf{Objective:} 
However, the scarcity of labeled samples and the class imbalance issue in vulnerability datasets present significant challenges for both Vulnerability Type Prediction (VTP) and Line-level Vulnerability Detection (LVD), especially for rare yet critical vulnerability types. Moreover, most existing studies treat VTP and LVD as independent tasks, overlooking their inherent correlation, which limits the potential to leverage shared semantic patterns across tasks.

\textbf{Methods:} To address these limitations, we propose a unified approach that integrates Embedding-Layer Driven Adversarial Training (EDAT) with Multi-task Learning (MTL). Specifically, EDAT enhances model robustness by introducing adversarial perturbations to identifier embeddings, guided by semantic importance. Meanwhile, MTL improves overall performance by leveraging shared representations and inter-task correlations between VTP and LVD.

\textbf{Results:} Extensive experiments demonstrate that our proposed approach outperforms state-of-the-art baselines on both VTP and LVD tasks. For VTP, it yields notable improvements in accuracy, precision, recall, and F1-score, particularly in identifying rare vulnerability types. Similarly, for LVD, our approach enhances line-level detection accuracy while significantly reducing false positives.

\textbf{Conclusion:} Our study demonstrates that combining EDAT with MTL provides a unified solution that improves performance on both tasks and warrants further investigation.

\end{abstract}

\begin{keyword}
    Vulnerability type prediction; 
    Line-level vulnerability detection;
    Adversarial training;
    Identifier embedding;
    Data augmentation;
    Multi-task learning
\end{keyword}

\end{frontmatter}

\section{Introduction}
\label{sec:intro}

Software vulnerabilities have become an increasingly critical concern in modern software systems, as evidenced by the growing number of reported vulnerabilities and cyberattacks in recent years. According to the National Vulnerability Database (NVD), more than 28,830 new vulnerabilities were disclosed in 2023~\cite{alfasi2024vulnscopper}. In early 2024, the NVD~\cite{nvd2024announcement} announced that it could no longer enrich newly published Common Vulnerabilities and Exposures (CVEs) due to the overwhelming volume, significantly hindering global cybersecurity efforts. These vulnerabilities not only threaten the security and reliability of software applications but also impose substantial financial and operational burdens on organizations and developers responsible for maintaining secure codebases~\cite{jiarpakdee2020empirical}. This escalating situation underscores the urgent need for effective, reliable, and automated techniques for detecting and understanding software vulnerabilities.

Two critical tasks in software vulnerability detection are Vulnerability Type Prediction (VTP) and Line-level Vulnerability Detection (LVD). Specifically, VTP focuses on classifying vulnerabilities into predefined categories, such as buffer overflow or SQL injection, while LVD aims to pinpoint the exact lines of code that contain security flaws. VTP is generally approached as a classification task, whereas LVD operates at a finer granularity, identifying vulnerable code at the line level to enable more precise and efficient remediation~\cite{hin2022linevd}.

Despite notable progress in vulnerability detection research, several limitations remain in existing studies.
First, traditional data augmentation strategies~\cite{rahman2023data,xie2020unsupervised} often rely on superficial perturbations, such as token-level replacements or character substitutions, which may compromise the syntactic or semantic correctness of the code. These approaches struggle to generalize across structurally diverse vulnerabilities, especially those involving complex program logic.
Second, prior works~\cite{menzies2011local,nam2013transfer} typically treat Vulnerability Type Prediction (VTP) and Line-level Vulnerability Detection (LVD) as separate tasks. This isolation overlooks their inherent correlation, thereby limiting generalization and resulting in suboptimal detection performance in real-world scenarios.

To address the aforementioned limitations, we propose a unified framework called Embedding-Layer Driven Adversarial Training and Multi-task Learning (EDAT-MTL). This framework integrates two complementary modules aimed at improving both the robustness and generalization of vulnerability detection models. The first module enhances the quality and semantic consistency of adversarial data augmentation, while the second leverages inter-task correlations through shared multi-task learning. By combining these modules, EDAT-MTL enables a unified training strategy that simultaneously boosts the performance of both VTP and LVD tasks.

To address the challenge of preserving code semantics during augmentation, we introduce the Embedding-Layer Driven Adversarial Training (EDAT) module. EDAT enhances model robustness by injecting perturbations into the embeddings of code identifiers, guided by attention-based signals, where the perturbation strength is proportional to each token's semantic importance. The EDAT workflow consists of five key steps: (1) input embedding initialization, (2) identifier extraction and perturbation target selection, (3) adversarial perturbation optimization using multi-step Projected Gradient Descent (PGD), (4) enforcement of code semantics preservation via Abstract Syntax Tree (AST) and Program Dependency Graph (PDG) constraints, and (5) generation of final adversarial examples. This module ensures that the introduced perturbations preserve both syntactic validity and functional correctness, while improving the model’s sensitivity to semantically significant identifier variations.

To mitigate the lack of interaction between the VTP and LVD tasks, we propose the Multi-task Learning (MLT) module. MLT enables joint optimization by employing a shared encoder to generate contextual embeddings for both tasks. The framework consists of three coordinated components: (1) CWE-guided joint encoding, which leverages Program Dependency Graph (PDG) structures to enhance task-relevant semantics; (2) task-aware attention fusion, which aligns the prediction heads by injecting class-specific attention signals; and (3) an uncertainty-aware consistency validation mechanism, which adaptively balances task-specific losses based on entropy and prediction variance. Together, these components improve type-level discrimination and line-level detection accuracy, while also reducing overfitting and enhancing generalization, particularly in long-tail vulnerability scenarios.

To evaluate the effectiveness of the proposed EDAT-MLT framework, we conduct comprehensive comparisons against state-of-the-art baselines for both the VTP and LVD tasks. For VTP, we compare with LIVABLE~\cite{wen2024livable} and VulExplainer~\cite{fu2023vulexplainer}, while for LVD, we include LineVD~\cite{hin2022linevd} and LineVul~\cite{fu2022linevul} as baselines.
Experimental results demonstrate that EDAT-MLT consistently outperforms all baselines. Specifically, on the VTP task, our model achieves an F1-score of 0.7281, surpassing LIVABLE (0.5469) and VulExplainer (0.5095) by 33.2\% and 42.9\%, respectively. On the LVD task, EDAT-MLT achieves a Top-5 Accuracy of 0.6047 on CodeBERT, representing a 157.4\% improvement over LineVD and a 56.7\% improvement over LineVul.

To validate the design rationale of the EDAT module, we conduct a targeted ablation study. For the VTP task, EDAT enhances the model’s generalization capability, particularly for rare and long-tail vulnerability types. For example, on CodeT5, incorporating EDAT increases the F1-score from 62.47\% to 74.76\%, with a corresponding recall improvement from 66.85\% to 76.12\%. For the LVD task, EDAT effectively reduces false positives and improves line-level localization precision. Specifically, on CodeBERT, EDAT raises Recall@20\% LOC from 0.5582 to 0.6519, while reducing the Initial False Alarm (IFA) from 6.2145 to 2.7913.

To assess the effectiveness of the MLT module, we conduct a targeted ablation study by comparing single-task and multi-task variants of our approach using three different pre-trained code models as backbones. For the VTP task, MLT consistently improves performance. Specifically, the F1-score increases from 0.4857 to 0.5965 on CodeBERT, from 0.5185 to 0.6258 on GraphCodeBERT, and from 0.5456 to 0.6427 on CodeT5, corresponding to relative improvements of 22.8\%, 20.7\%, and 17.8\%, respectively.
For the LVD task, MLT similarly enhances localization performance across all pre-trained models. For example, on CodeT5, Recall@20\% LOC increases from 0.3616 to 0.6278, while the Initial False Alarm decreases from 11.8560 to 5.2896.

Our study proposes a unified solution to address key limitations in prior vulnerability detection research, specifically focusing on challenges related to data augmentation and the exploitation of task correlations between VTP and LVD. By introducing the EDAT-MLT framework and conducting evaluations on a dataset collected from real-world vulnerabilities, we demonstrate that a unified learning strategy not only enhances model robustness through effective data augmentation but also facilitates synergistic feature extraction across tasks. This approach offers a promising direction for improving both vulnerability type prediction and line-level detection performance and represents a valuable avenue for future research in the field of automated vulnerability analysis.

In summary, the main contributions of our study can be summarized as follows:

\begin{itemize}
    \item We propose a unified framework, EDAT-MLT, which integrates Embedding-Layer Driven Adversarial Training and Multi-task Learning to jointly enhance vulnerability type prediction and line-level vulnerability detection.
    
    \item We design the EDAT module to inject semantically-aware perturbations into identifier embeddings while preserving program logic, thereby enhancing model robustness, particularly against rare and long-tail vulnerabilities.
    
    \item We employ the MLT module to facilitate cross-task knowledge sharing through a shared encoder, thereby improving the performance of both vulnerability type prediction and line-level detection.
    
    \item Extensive experiments on the popular vulnerability detection dataset demonstrate that EDAT-MLT consistently outperforms state-of-the-art baselines. Furthermore, ablation studies confirm the individual effectiveness of both the EDAT and MLT modules.
\end{itemize}

\textbf{Open Science.} 
To support the open science community, we have made the dataset, scripts (covering data preprocessing, model training, and performance evaluation), and experimental results publicly available in a GitHub repository: 

https://github.com/Karelye/EDAT-MLT

\textbf{Paper Organization.}
Section~\ref{sec:Background} introduces the background of our study.
Section~\ref{sec:Motivation} presents the overall framework of our proposed approach along with detailed descriptions of each module.
Section~\ref{sec:Experiment} describes the experimental setup, followed by Section~\ref{sec:Result}, which analyzes the experimental results.
Section~\ref{sec:Discussions} discusses the impact of hyperparameter settings and potential threats to validity.
Section~\ref{sec:Related} reviews related work and highlights the novelty of our study.
Finally, Section~\ref{sec:Conclusion} concludes the paper and outlines potential directions for future research.

\section{Background}
\label{sec:Background}

In this section, we introduce the background of our study, focusing on four key areas: vulnerability type prediction, line-level vulnerability detection, data augmentation, and multi-task learning.

\subsection{Vulnerability Type Prediction}

Vulnerability type prediction~\cite{cito2023expert,jiarpakdee2020empirical,jiarpakdee2021practitioners,khanan2020jitbot,liu2022explainable,pornprasit2021pyexplainer,rajapaksha2021sqaplanner,tantithamthavorn2023explainable,tantithamthavorn2021actionable} is a fundamental task in secure software engineering, which aims to automatically identify the type of vulnerability in source code, typically using the Common Weakness Enumeration (CWE) taxonomy. 
Accurate VTP enables effective risk prioritization, patch planning, and compliance reporting. Traditional VTP methods~\cite{jiarpakdee2020empirical,khanan2020jitbot,pornprasit2021pyexplainer} leverage code features by static or dynamic code analysis and train classifiers to map code snippets to predefined CWE types. Recent advances have incorporated neural code representations based on abstract syntax trees~\cite{sun2023ast}, control/data-flow graphs, and pre-trained language models, to capture structural semantics and improve prediction accuracy. These advances have enhanced vulnerability type prediction performance on multiple public datasets, further establishing VTP as a core component of automated vulnerability analysis pipelines.

More recently, researchers continued to explore advanced strategies to improve the effectiveness and interpretability of vulnerability type prediction. Building upon the foundation of neural code representations and pre-trained models, Jiarpakdee et al.~\cite{jiarpakdee2020empirical} introduced an automated approach for explaining vulnerability prediction models by quantifying feature influence through permutation-based analysis. Their empirical evaluation demonstrated the method's applicability across different datasets and classifiers. Similarly, Tantithamthavorn et al.~\cite{tantithamthavorn2021actionable} proposed an actionable analytics framework that integrates explainable AI with rule-based and instance-specific guidance. Through comprehensive user studies, they validated the practical benefits of their approach in enhancing developers’ understanding and trust in automated vulnerability classification results. These efforts highlight a growing emphasis on improving classification accuracy and delivering interpretable and generalizable solutions in real-world software development and maintenance.

\subsection{Line-level Vulnerability Detection}

Line-level vulnerability detection focuses on identifying specific lines of code that are vulnerable to security flaws, enabling more precise remediation efforts compared to traditional software vulnerability classification. It typically operates by analyzing code-line granularity through semantic pattern recognition, which is achieved by parsing Abstract Syntax Trees or code embeddings using graph neural networks and attention mechanisms~\cite{fu2022linevul, hin2022linevd, nguyen2021information, he2025telecontext, wattanakriengkrai2020predicting}. This approach excels at identifying lines that contain unsafe operations, such as improper index access or unchecked inputs, which are crucial for prioritized auditing. 

To further improve LVD performance, Hin et al.~\cite{hin2022linevd} propose LineVD, a novel graph-based model that integrates multi-view representations of syntax, semantics, and control flow to capture structural dependencies and enhance statement-level detection accuracy. Additionally, Fu et al.~\cite{fu2022linevul} propose LineVul, a transformer-based model that encodes both syntactic and semantic contextual features, achieving state-of-the-art performance on multiple real-world datasets. These two studies can highlight the growing emphasis on fine-grained line-level vulnerability detection, which complements type-level analysis by enabling precise and actionable detection of insecure code regions.

\subsection{Data Augmentation}

Adversarial training has become a widely adopted data augmentation strategy to improve model robustness, particularly in tasks requiring high semantic fidelity, such as vulnerability detection. By introducing small perturbations near decision boundaries, models learn to generalize beyond seen data~\cite{goodfellow2014explaining, li2018learning}. In code analysis, this helps capture semantically consistent but structurally diverse representations. However, traditional adversarial approaches on software engineering tasks often operate at the character level~\cite{dey2024semantic}, modifying tokens or syntax in ways that compromise program validity. These methods, effective in natural language tasks~\cite{rahman2023data, xie2020unsupervised}, struggle to preserve code semantics and structure, limiting their applicability to software-related tasks.

To address the above limitations, recent studies shift towards embedding-level perturbations specifically targeting identifier embeddings~\cite{yang2022natural, zhang2022towards}. By injecting gradient-driven noise into these identifier embeddings (such as those for variables and functions), this strategy enables automatic discovery of invariant transformations, dynamic adaptation during training, and targeted reinforcement of minority CWE-ID classes. This approach of augmenting identifier embeddings maintains syntactic correctness while introducing controlled semantic variation, offering a principled solution for robust vulnerability classification with the class imbalance issue.

\subsection{Multi-task Learning}

Multi-task learning has emerged as an effective paradigm for improving generalizability in software engineering tasks by jointly optimizing multiple related objectives. Unlike single-task models, which are often limited by data sparsity and redundant parameter learning, MTL leverages shared representations to transfer knowledge across tasks~\cite{ni2019multitask, huang2024multi}. By jointly training on related tasks, such as vulnerability type prediction and line-level vulnerability detection, MTL frameworks can help to reduce overfitting and improve model robustness, especially under limited labeled data scenarios~\cite{yang2024multi}.

Technically, MTL architectures adopt shared and task-specific components to enable knowledge sharing while preserving task uniqueness. Early studies decouple shared and private layers to balance information transfer~\cite{ni2019multitask}, while recent approaches utilize attention mechanisms and multi-view fusion strategies to model complex cross-task dependencies~\cite{huang2024multi, yang2024multi}. For example, COMPDIRECT and JIT-Smart demonstrate the benefits of aligning type prediction, line-level detection, and repair in unified frameworks~\cite{ni2023unifying, chen2024ijit}. Inspired by these studies, we extend MTL to software vulnerability analysis by jointly modeling vulnerability type prediction (i.e., CWE-ID classification) and line-level detection. This integration enables bidirectional supervision: global type semantics guide spatial focus, while localized cues refine type decisions. Our design offers a unified architecture capable of enhancing vulnerability understanding across coarse and fine granularities.

 \section{Research Motivation and Our Proposed Approach}
\label{sec:Motivation}

\subsection{Research Motivation}
\label{subsec:Research Motivation}

In this subsection, we analyze the main limitations of previous studies, which serve as the motivation for our work.

\textbf{Semantic Misalignment in Adversarial Augmentation.}
Traditional rule-based and character-level adversarial methods perturb code tokens or characters without preserving syntactic or functional integrity~\cite{rahman2023data, xie2020unsupervised, goodfellow2014explaining, li2018learning}. Such approaches often disrupt program logic by introducing invalid token substitutions and exhibit limited effectiveness in vulnerability detection. Moreover, they primarily target superficial variations, like random renaming or formatting changes, while failing to capture deeper structural patterns relevant to vulnerabilities. Consequently, models trained with these augmentations tend to overfit to lexical features and struggle to generalize to inputs that are semantically similar but structurally distinct.

\textbf{Ignoring Structural Interdependence Between Tasks.}
Traditional approaches in software vulnerability detection often treat vulnerability type prediction and line-level vulnerability detection as separate tasks~\cite{menzies2011local, nam2013transfer, turhan2009relative, zhou2018far}. This separation overlooks the intrinsic correlation between these tasks, ignoring the structural interdependence between vulnerability types and their specific manifestations in source code. For instance, buffer overflow vulnerabilities (CWE-787) typically stem from inadequate boundary checks in loops, while SQL injection vulnerabilities (CWE-89) often arise due to improper sanitization of user inputs in database queries. Such examples demonstrate that the semantics of vulnerability types are deeply intertwined with contextual and structural cues at the line level, underscoring the necessity for models capable of jointly reasoning about both aspects.

\subsection{ Our Proposed Approach}
\subsubsection{Framework}

\begin{figure*}[htbp] 
    \centering 
    \includegraphics[width=\textwidth]{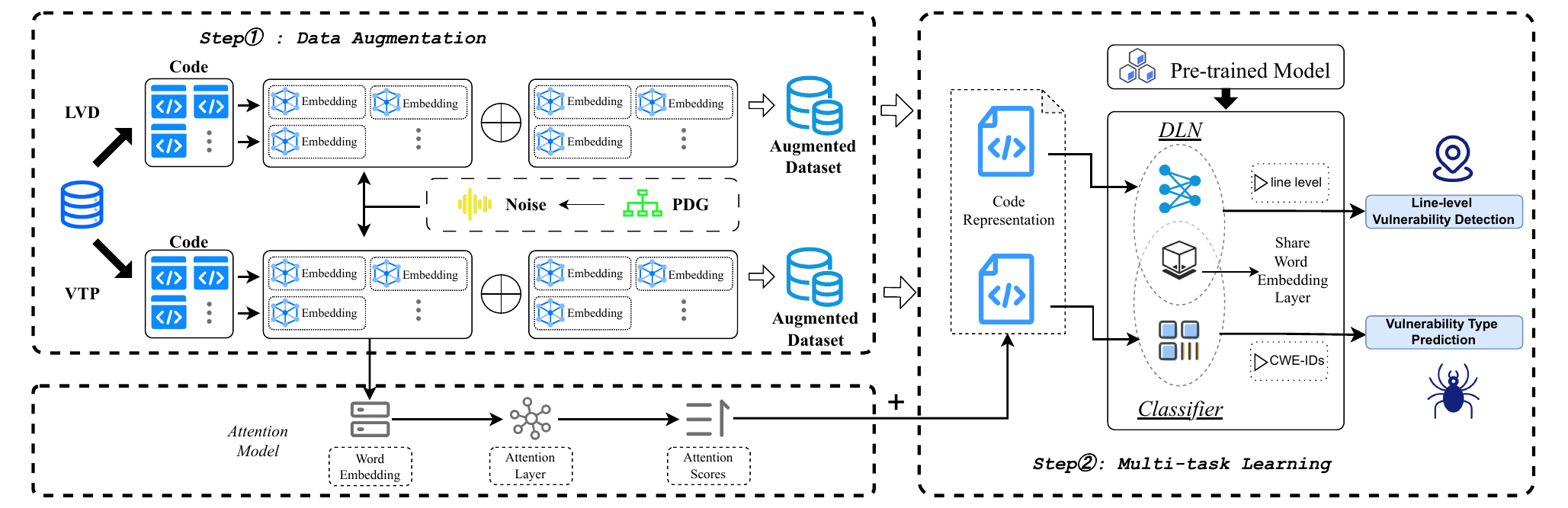}
\caption{An overview of our proposed approach EDAT-MLT} 
    \label{Fig:overview1} 
\end{figure*}

As illustrated in Figure~\ref{Fig:overview1}, our EDAT-MLT framework consists of two core modules. The first module, Embedding-Layer Driven Adversarial Training (EDAT), injects adversarial noise into code identifier embeddings guided by attention mechanisms. Syntactic parsing is used to identify non-critical targets, while semantic validation ensures functional integrity during perturbation optimization. The second module, Multi-task Learning (MTL), employs a shared encoder with frozen lower layers and fine-tuned upper layers to generate contextual embeddings. Attention mechanisms are leveraged to emphasize code regions associated with perturbed identifiers, enabling joint training for Vulnerability Type Prediction and Line-level Vulnerability Detection, and promoting feature alignment across tasks. The following subsections provide detailed descriptions of each module.

\subsubsection{Data Augmentation Module}

To address the first limitation discussed in Section~\ref{subsec:Research Motivation}, the data augmentation module utilizes Embedding-Layer Driven Adversarial Training (EDAT). EDAT enhances model robustness by generating adversarial perturbations on identifier embeddings while preserving code semantics. The process consists of five steps, as outlined in Algorithm~\ref{alg:EDAT}.

\begin{algorithm}[htbp]
\caption{Embedding-layer driven adversarial training}
\label{alg:EDAT}
\begin{algorithmic}[1] 
\item[] \hspace{-1.5em} \textbf{Inputs:}
\Statex \hspace{0.5em} $D \gets \{(x, y)\}$: training sets;
\Statex \hspace{0.5em} $K$: ascent steps;
\Statex \hspace{0.5em} $\mu$: adversarial learning rate;
\Statex \hspace{0.5em} $\varepsilon$: perturbation bound;
\Statex \hspace{0.5em} $N$: total number of training epochs\;
\Statex \hspace{0.5em} $m$: number of samples per mini-batch\;
\item[] \hspace{-1.5em} \textbf{Outputs:}
\Statex \hspace{0.5em} $\theta_c$: model $M_c$'s parameters;
\For{\text{epoch} $\gets 1, \dots, N$}
    \For{\text{mini-batch} $B \in D$} 
        \For{$i \gets 1, \dots, m$}
            \item[] // Compute initial loss for both models
            \State $p_a \gets M_a(x_i)$
            \State $p_c \gets M_c(x_i)$
            \State $\theta_a \gets \theta_a + \eta \delta L_{\theta_a}$
            \State \text{update} $\theta_c$
            \item[] // Extract identifiers and get embedding token\_ids
            \State $\text{target\_ids} \gets \text{ExtractIdentifiers}(x_i)$
            \For{$t \gets 1$ \text{to} $K$}
                \item[] // Compute adversarial perturbation
                \State $g_i \gets \nabla L(M_a(x_i + \delta), y_i)$
                \State $\delta_i \gets \delta_i + \mu g_i / ||g_i||$
            \EndFor
            \item[] // Add perturbation to embeddings of identifiers
            \State $\text{embeddings}[X] \gets \text{embeddings}[X] + \delta$
            \item[] // Compute adversarial loss after perturbation
            \State $p_a \gets M_a(x_i)$ 
            \State $p_c \gets M_c(x_i)$
            \State $\theta_a \gets \theta_a + \eta \delta L_{\theta_a}$
            \State \text{update} $\theta_c$ according to $\delta L_{\theta_c}$
        \EndFor
    \EndFor
\EndFor
\State \textbf{return} $\theta_c$
\end{algorithmic}
\end{algorithm}
 
\textbf{Step 1: Input Embedding Initialization.}
Given a training set $D = \{(x, y)\}$, we first tokenize the input code $x$ into $input\_ids$ and feed them into a pretrained encoder to generate token-level embeddings $E \in \mathbb{R}^{T \times d}$, where $T$ denotes the sequence length and $d$ is the embedding dimension. For vulnerability type prediction, the [CLS] token embedding is used. For line-level detection, line-wise embeddings are computed by average pooling over tokens within the same line.

\textbf{Step 2: Identifier Extraction and Perturbation Target Generation.}
We apply syntax parsing (e.g., Tree-sitter\footnote{\url{https://tree-sitter.github.io/tree-sitter/}}) to extract code identifiers such as variable and function names. Their token indices are recorded as the perturbation targets $X$. The embedding vector $e_i$ for each identifier is perturbed with Gaussian noise:

\begin{equation}
\delta^{(0)}_i \sim \mathcal{N}(0, \sigma^2 \cdot \alpha_i)
\end{equation}
where $\alpha_i$ is the attention-based importance score of identifier $i$, determining how much noise it receives. Identifiers with higher semantic significance (e.g., tainted variables) receive stronger perturbations.

\textbf{Step 3: Adversarial Perturbation Optimization via Multi-step Projected
Gradient Descent.}
To refine the perturbation $\delta_i$, we adopt multi-step PGD to maximize the model's classification loss while preserving code functionality:

\begin{equation}
\delta^{(t+1)}_i = \text{Clip}_\epsilon \left( \delta^{(t)}_i + \mu \cdot \frac{\nabla_{\delta} \mathcal{L}(M(e_i + \delta^{(t)}_i), y)}{ \| \nabla_{\delta} \mathcal{L} \|_2 } \right)
\end{equation}
where $\mu$ is the adversarial learning rate, $\epsilon$ is the perturbation bound, and $\mathcal{L}$ is the classification loss. The gradient is normalized to prevent excessive updates. This process generates semantically aligned but function-preserving adversarial samples.

\textbf{Step 4: Code Semantics Preservation via AST and PDG Constraints.}
To guarantee semantic consistency, only non-functional identifiers are selected via Abstract Syntax Tree (AST) filtering. We further validate perturbation legality using Program Dependency Graph (PDG), ensuring that renamed variables do not alter control or data dependencies (e.g., across def-use chains). Perturbations violating these constraints are discarded.

\textbf{Step 5: Adversarial Sample Generation.}
The final perturbed embeddings are added to the input embeddings and passed through the model. The model parameters $\theta$ are updated based on the adversarial loss, defined as a combination of cross-entropy loss for type identification, focal loss for line-level detection, and KL divergence to enforce perturbation invariance. This loss function ensures that the model learns to focus on semantically meaningful signals while being robust to identifier renaming, formatting changes, and rare pattern variations.

\subsubsection{Multi-task Learning Module}

To address the second limitation discussed in Section~\ref{subsec:Research Motivation}, we design a unified Multi-task Learning (MTL) module. Traditional approaches often treat these tasks independently, neglecting the semantic interdependencies inherent in real-world vulnerability contexts. Such separation impairs model generalization and interpretability, particularly given the scarcity of labeled data and the presence of class imbalance. The MTL module bridges this gap by leveraging shared representations and enabling mutual constraint propagation between tasks.

Our MTL framework consists of three tightly integrated components, each targeting a specific challenge in multi-task vulnerability analysis:

\textbf{(1) CWE-Semantic Guided Joint Encoding.} 
To ensure contextual alignment across tasks, the unified encoder processes input tokens using Program Dependency Graph (PDG)-based structural masks $M_{PDG}$ to filter out irrelevant interactions. This design selectively amplifies code features pertinent to both semantic type identification and spatial line-level detection, reducing interference from unrelated patterns. Additionally, it enhances the interpretability of perturbations generated by the EDAT module by concentrating updates on semantically significant regions.

\textbf{(2) Cross-Task Attention Fusion.} 
While type identification and line-level detection differ in granularity, their correlation can be effectively captured through a task-aware attention mechanism. Specifically, each task’s attention head incorporates the prediction embedding from the other task as an additional query bias, thereby forming joint attention distributions over token representations. This design facilitates knowledge transfer: classification predictions reinforce attention on localizable contexts, while localized code regions provide valuable cues for semantic understanding. Such coupling not only reduces false positives but also promotes consistency across tasks.

\textbf{(3) Uncertainty-Aware Consistency Validation.} To ensure training stability under cross-task supervision, we introduce an adaptive loss weighting scheme based on task uncertainty. The combined loss is defined as:

\begin{equation}
\mathcal{L}_{\text{joint}} = \lambda \mathcal{L}_{cls} + (1 - \lambda) \mathcal{L}_{loc}
\end{equation}
where the weighting factor $\lambda$ is dynamically adjusted using entropy and prediction variance:

\begin{equation}
\lambda = \frac{1}{1 + e^{-\text{Var}(p_{loc}) - \text{Entropy}(p_{cls})}}
\end{equation}

This strategy encourages the model to prioritize line-level detection signals when the confidence of type identification is low, and vice versa, thereby enhancing robustness in handling long-tail categories and ambiguous inputs.


\section{Experimental Setup}
\label{sec:Experiment}

\subsection{Research Questions}

To systematically assess the effectiveness of our proposed approach, we design three research questions (RQs) that guide our empirical evaluation.

\textbf{RQ1: Can our proposed approach outperform the baseline methods on vulnerability type prediction and line-level vulnerability detection?}

\textbf{Design Motivation.}
Previous vulnerability detection studies~\cite{hin2022linevd,wen2024livable} often treat Vulnerability Type Prediction and Line-level Vulnerability Detection as independent tasks, resulting in disjoint inference and reduced reasoning coherence.
For the VTP task, we compare our approach with two state-of-the-art baselines: VulExplainer~\cite{fu2023vulexplainer}, which addresses class imbalance through hierarchical knowledge distillation, and LIVABLE~\cite{wen2024livable}, which leverages graph neural networks and bidirectional LSTMs to capture structural and semantic features.
For the LVD task, we consider LineVD~\cite{hin2022linevd}, a graph-based model that uses attention to model inter-statement context, and LineVul~\cite{fu2022linevul}, a Transformer based method that enables direct line-level detection without relying on explicit program analysis.
These baselines treat VTP and LVD separately, often overlooking their intrinsic correlation. Moreover, challenges such as data scarcity for rare vulnerability types and insufficient context-aware modeling further limit their effectiveness.
To address these limitations, we propose the EDAT-MLT framework, which integrates embedding-layer adversarial training with multi-task learning to enhance both robustness and cross-task synergy. In this RQ, we want to evaluate our approach on the Big-Vul dataset~\cite{fan2020ac} to demonstrate its effectiveness.

\textbf{RQ2: Can embedding-layer driven data augmentation enhance performance in vulnerability type prediction and line-level vulnerability detection?}

\textbf{Design Motivation.} 
Prior studies~\cite{goodfellow2014explaining,li2018learning} typically introduce perturbations at the character or token level, frequently resulting in syntactic violations or semantic inconsistencies. Such perturbations may obscure vulnerability-relevant patterns and reduce the effectiveness of augmented training data.
To mitigate these issues, we propose an Embedding-Layer Driven Adversarial Training (EDAT) module, which perturbs identifier embeddings under syntax- and semantics-aware constraints. This design ensures that the generated adversarial samples preserve both syntactic correctness and semantic coherence, thereby maintaining the functional integrity of the original code. As a result, EDAT improves model generalization and robustness in both Vulnerability Type Prediction and Line-level Vulnerability Detection.
In this RQ, we aim to conduct an ablation study to evaluate the effectiveness of our designed EDAT module.

\textbf{RQ3: Can multi-task learning improve the performance of vulnerability type prediction and line-level vulnerability detection?}

\textbf{Design Motivation.}
Traditional single-task models treat VTP and LVD as independent tasks, inherently limiting their ability to exploit the shared semantic and structural patterns that exist between them. This artificial decoupling hinders potential knowledge transfer. For example, using type-specific code patterns (e.g., buffer overflow signatures) to guide fine-grained localization, or leveraging line-level vulnerability distributions to refine type classification.
To address this limitation, we adopt an MTL framework with a shared encoder, enabling joint optimization across VTP and LVD. This design encourages complementary feature extraction: VTP benefits from context representations enriched by line-level detection signals, while LVD gains discriminative insights from high-level vulnerability semantics associated with specific types. Such cross-task interaction fosters a holistic understanding of software vulnerabilities.
In this RQ, we aim to empirically evaluate whether MTL leads to superior performance compared to state-of-the-art single-task baselines. Specifically, we investigate whether coordinated multi-task learning enables better generalization and robustness by capturing inter-task dependencies that isolated models fail to leverage.

\subsection{Experimental Subjects}

In this study, we adopt the Big-Vul dataset~\cite{fan2020ac}, a large-scale C/C++ corpus that provides comprehensive vulnerability annotations at both the function level and the line level. This dual-granularity annotation makes Big-Vul particularly well-suited as a benchmark for evaluating the two core tasks of our framework: VTP and LVD. The dataset is constructed from 348 real-world open-source repositories and covers 3,754 CVEs, ensuring high authenticity and diversity. A detailed statistical summary is presented in Table~\ref{tab:bigvul_stats_compact}.

For the VTP task, we leverage the function-level labels along with their associated Common Weakness Enumeration (CWE) annotations. The presence of 91 distinct CWE types presents a realistic and challenging setting for evaluating a model's capacity to distinguish between diverse and semantically nuanced vulnerability classes. 

For the LVD task, we utilize the fine-grained line-level annotations available in Big-Vul. Specifically, the dataset includes 44,603 labeled vulnerable lines distributed across 11,823 vulnerable functions, offering precise ground truth for both training and evaluation. The sparse distribution of vulnerable lines (with a median of only 7.0\%) can reflect real-world software development scenarios and provides a rigorous evaluation of a model's vulnerability type identification and line-level detection.

\begin{table}[htbp]
\centering
\scriptsize
\caption{Descriptive Statistics of the Big-Vul Dataset}
\label{tab:bigvul_stats_compact}
\setlength{\tabcolsep}{4pt} 
\resizebox{\linewidth}{!}{
\begin{tabular}{lll}
\toprule
\textbf{Category} & \textbf{Metric} & \textbf{Value / Description} \\
\midrule
\multirow{5}{*}{Dataset Metadata} 
 & Number of open-source repositories & 348 \\
 & Number of CVEs & 3,754 \\
 & Number of CWE types & 91 \\
 & Number of features & 21 \\
\midrule
\multirow{3}{*}{Function-Level Statistics} 
 & Vulnerable vs. non-vulnerable functions & 11,823 / 253,096 \\
 & Total number of C/C++ functions & 188,636 \\
 & Proportion of vulnerable functions & 5.7\% \\
\midrule
\multirow{3}{*}{Line-of-Code (LOC) Level} 
 & Total lines of code (LOC) & 5,060,449 \\
 & Number of vulnerable lines of code & 44,603 \\
 & Proportion of vulnerable lines of code & 0.88\% \\
\midrule
Vulnerability Distribution & \begin{tabular}[t]{@{}l@{}}Proportion of vulnerable LOC \\ within vulnerable functions\end{tabular} & \begin{tabular}[t]{@{}l@{}}Median: 7\% \\ 1st Quartile: 2.5\% \\ 3rd Quartile: 20\%\end{tabular} \\
\bottomrule
\end{tabular}
}
\end{table}

\subsection{Performance Metrics}

To evaluate the performance of our proposed approach, we employ a comprehensive suite of performance metrics for both vulnerability type prediction and line-level vulnerability detection tasks.

\subsubsection{Vulnerability Type Prediction}
To evaluate the performance of the VTP task, we consider the following performance metrics.

\textbf{Accuracy.}
Accuracy measures the proportion of correctly classified samples among all instances. It is defined as:
\begin{equation}
\text{Accuracy} = \frac{TP + TN}{TP + TN + FP + FN}
\end{equation}
where $TP$ and $TN$ represent the number of true positive and true negative predictions, respectively. $FP$ denotes false positives, and $FN$ denotes false negatives. This metric reflects the overall classification correctness.

\textbf{Precision.}
Precision quantifies the exactness of positive predictions for a specific class $c$ and is computed by:
\begin{equation}
\text{Precision}_c = \frac{TP_c}{TP_c + FP_c}
\end{equation}
where $TP_c$ refers to the number of class-$c$ instances correctly predicted as positive, while $FP_c$ denotes the number of instances incorrectly predicted as class $c$. High precision indicates a low false positive rate.

\textbf{Recall.}
Recall, also known as sensitivity, assesses the model's ability to identify all relevant instances of a class $c$:
\begin{equation}
\text{Recall}_c = \frac{TP_c}{TP_c + FN_c}
\end{equation}
where $TP_c$ is the number of correctly identified class-$c$ instances, and $FN_c$ denotes the class-$c$ instances missed by the model. This metric is particularly important in scenarios where missing vulnerabilities is costly.

\textbf{F1-Score.}
The F1-score provides a harmonic mean between precision and recall, offering a balanced measure:
\begin{equation}
\text{F1}_c = \frac{2 \times \text{Precision}_c \times \text{Recall}_c}{\text{Precision}_c + \text{Recall}_c}
\end{equation}
It is sensitive to class imbalance and is useful when both false positives and false negatives are consequential.

\subsubsection{Line-level Vulnerability Detection}
To evaluate the performance of the LVD task, we consider the following performance metrics.

\textbf{Top-5/10 Accuracy.} Top-5/10 Accuracy is a ranking-based metric that evaluates whether the true vulnerable line appears among the model’s top-k predictions, where $k \in \{5, 10\}$. A higher Top-5/10 Accuracy indicates that the model ranks the actual vulnerable lines higher, which facilitates efficient triage by developers. The metric is formally defined as:

\begin{equation}
\text{Top-}k\ \text{Acc} = \mathbb{I}(l \in \mathcal{V} \cap \{1, \ldots, k\}), \quad k = 5\ \text{or}\ 10,
\label{eq:topkacc}
\end{equation}
where $\mathcal{V}$ denotes the set of ground-truth vulnerable lines, $l$ is the line index predicted by the model ranked by confidence score, and $\mathbb{I}$ is the indicator function that returns 1 if any of the top-$k$ predictions intersect with the true vulnerable lines, and 0 otherwise.

\textbf{Recall@20\% LOC.}
This metric quantifies how many true vulnerable lines are identified when only the top 20\% of lines in a file (sorted by model confidence) are inspected:
\begin{equation}
\text{Recall@20\% LOC} = \frac{|\mathcal{V}_{\text{detected}}|}{|\mathcal{V}_{\text{total}}|}
\end{equation}
where $\mathcal{V}_{\text{detected}}$ denotes the number of ground-truth vulnerable lines ranked within the top 20\% LOC by model confidence, and $\mathcal{V}_{\text{total}}$ is the total number of ground-truth vulnerable lines. A higher value indicates stronger detection capability under limited inspection effort.

\textbf{Effort@20\% Recall.}
Effort@20\% Recall measures the minimum proportion of lines that must be inspected to detect 20\% of vulnerabilities:
\begin{equation}
\text{Effort@20\% Recall} = \frac{\text{Rank}(l_q)}{N}, \quad q = \left\lfloor 0.2 \times |\mathcal{V}| \right\rfloor
\end{equation}
where $N$ is the total number of lines, and $\text{Rank}(l_q)$ gives the index of the $q$th correctly predicted vulnerable line in the descending confidence-ranked list.

\textbf{Initial False Alarm (IFA).}
IFA measures how many non-vulnerable lines must be reviewed before encountering the first true vulnerable line:
\begin{equation}
\text{IFA} = \min\{k \,|\, \text{Line}_k \in \mathcal{V}\} - 1
\end{equation}
where $\mathcal{V}$ is the set of true vulnerable lines, $\text{Line}_k$ denotes the $k$th line in the descending prediction list, and the operator finds the earliest position where a true vulnerable line is located.

\subsection{Baselines}

In this subsection, we introduce state-of-the-art baselines for each task used to compare the performance of our proposed approach.
 
\subsubsection{Vulnerability Type Prediction}

For this task, we select VulExplainer and LIVABLE as the baselines.

\begin{itemize}
    \item \textbf{VulExplainer.} Fu et al.~\cite{fu2023vulexplainer} propose VulExplainer, a Transformer-based hierarchical distillation framework for vulnerability classification. To address the long-tailed CWE-ID distribution, it first groups vulnerability types into balanced subgroups via CWE abstraction hierarchies. Multiple TextCNN teachers are trained on these subgroups, and a Transformer-based student is distilled to generalize across all types. A distillation token is introduced to fuse teacher knowledge without modifying Transformer architectures, enabling compatibility with models like CodeBERT and GraphCodeBERT.

    \item \textbf{LIVABLE.} Wen et al.~\cite{wen2024livable} propose LIVABLE, a framework for long-tailed vulnerability classification that tackles two key challenges: over-smoothing in GNNs and data imbalance. It consists of two modules: (1) a representation learning module that uses differentiated propagation GNNs to mitigate over-smoothing and integrates a sequence-to-sequence model for semantic understanding; and (2) an adaptive re-weighting module that dynamically balances learning between head and tail classes via a dual-branch objective. This design improves the representation of rare types and enhances structural and semantic feature learning under long-tailed distributions.
\end{itemize}

\subsubsection{Line-level vulnerability detection}

For this task, we select LineVD and LineVul as the baselines.

\begin{itemize}
    \item \textbf{LineVD.} Hin et al.~\cite{hin2022linevd} propose LineVD, a statement-level vulnerability detection framework that formulates the task as a node classification problem over program dependency graphs. LineVD integrates token-level code semantics encoded by CodeBERT with statement-level control and data dependencies captured via Graph Attention Networks (GATs) to enhance granularity and interpretability. It uniquely fuses function-level and statement-level representations to jointly optimize prediction, enabling more precise localization of vulnerable lines. This design addresses the limitations of coarse-grained predictions in prior models and improves detection performance under realistic, imbalanced settings.

    \item \textbf{LineVul.} Fu et al.~\cite{fu2022linevul} propose LineVul, a Transformer-based model for fine-grained vulnerability prediction at the line level. To overcome limitations of RNN-based architectures and project-specific training data in prior work, LineVul leverages the pre-trained CodeBERT to capture long-range dependencies and semantic features. It uses attention-based line scoring mechanisms to localize vulnerable lines directly, without the need for explicit graph modeling or program analysis. This architecture enables LineVul to achieve more accurate and cost-effective line-level vulnerability detection, especially in large-scale and heterogeneous codebases.

\end{itemize}

\subsection{Code Pre-Trained Models}

To enhance the model's capability in capturing both syntactic structures and semantic patterns in source code, we adopt three widely used code pre-trained models as the backbone models of our proposed approach: CodeBERT~\cite{feng2020codebert}, GraphCodeBERT~\cite{guo2020graphcodebert}, and CodeT5~\cite{wang2021codet5}. These models are chosen for their promising effectiveness in various software engineering tasks~\cite{chen2024automatic,yang2024automatic,liu2023automated,yang2024important}, especially in vulnerability detection~\cite{li2023survey,zhu2022discovering}. 

\begin{itemize}
    \item \textbf{CodeBERT.}
    CodeBERT~\cite{feng2020codebert} is a bimodal pre-trained transformer model jointly trained on natural language and programming language pairs. It adopts a RoBERTa-like architecture and is trained using a masked language modeling objective and replaced token detection across six programming languages. In your study, CodeBERT serves as a foundational encoder that captures the textual and syntactic semantics of source code, making it suitable for global vulnerability classification tasks. However, due to its purely token-based design, it shows limited effectiveness in modeling fine-grained structural dependencies, which can hinder precise line-level detection.

    \item \textbf{GraphCodeBERT.}
    GraphCodeBERT~\cite{guo2020graphcodebert} extends CodeBERT by incorporating data flow graphs to model the relationships between variables in code. This hybrid approach enables the model to capture both sequential and structural information, particularly control and data dependencies. According to prior studies~\cite{li2023survey,zhou2023comprehensive}, GraphCodeBERT can enhance the model’s sensitivity to vulnerability-related variable interactions and is especially effective in LVD tasks. Its attention mechanism can distinguish semantically critical identifiers. Furthermore, its architecture's receptiveness to embedding-level manipulations makes it well-suited for adversarial perturbations targeting identifier embeddings.

    \item \textbf{CodeT5.}
    CodeT5~\cite{wang2021codet5} is a sequence-to-sequence Transformer model pre-trained on various programming-related tasks, including code summarization, translation, and generation. It utilizes a unified encoder-decoder architecture and is trained with denoising and span prediction objectives. In this research, CodeT5's generative pre-training provides superior adaptability for handling adversarial perturbations, particularly those on identifier embeddings. Its hybrid attention mechanisms further leverage this by enabling effective learning from the resulting semantically rich, embedding-level perturbed samples, which enhances both type identification and line-level detection capabilities.
\end{itemize}

The input data for our proposed model is organized along three dimensions: 
$L_c$, representing the total number of tokens for classification tasks; 
$N_l$, denoting the number of code lines involved in line-level tasks; and 
$N_t$, indicating the number of tokens per line for line-level tasks. Since the neural network is trained with mini-batches, fixed values must be set for these dimensions. If an input instance’s length is shorter than the fixed size along any dimension, padding is applied; if it exceeds the fixed size, truncation is performed.

We performed a statistical analysis of the distributions for the three input dimensions ($L_c$, $N_l$, and $N_t$) across the dataset. This analysis included the computation of key descriptive statistics: mean, standard deviation, minimum, and maximum values for each dimension. Based on these results, and to ensure comprehensive input coverage while maintaining computational efficiency, we set the fixed values as follows: $L_c = 512$ (code length for classification), $N_l = 256$ (number of lines for line-level tasks), and $N_t = 64$ (tokens per line for line-level tasks). These values were selected to ensure that more than 95\% of the data falls within the specified bounds, thus achieving a practical balance between data preservation and model efficiency.

\begin{table}[H] 
\centering
\scriptsize 
\caption{statistical analysis of the distributions for the three input dimensions} 
\setlength{\tabcolsep}{3pt} 

\resizebox{\linewidth}{!}{%
\begin{tabular}{@{}l c c c c c@{}} 
\hline
\textbf{Statistic} & \textbf{Coverage Ratio} & \textbf{Min} & \textbf{Max} & \textbf{Mean} & \textbf{Std} \\
\hline
Code length (train)  & 95.8\% & 2 & 3231 & 312.2& 396.5\\ \hline
Code lines (train)   & 95.8\% & 1 & 6092 & 55.5& 141.4\\ \hline
Code tokens (train)  & 99.8\% & 1 & 570  & 14.1& 9.3\\ \hline
Code length (valid)  & 95.1\% & 2 & 2686 & 323.5& 406.5\\ \hline
Code lines (valid)   & 97.0\% & 1 & 2515 & 47.4& 118.6\\ \hline
Code tokens (valid)  & 99.8\% & 1 & 635  & 15.1& 9.3\\ \hline
Code length (test)   & 96.6\% & 2 & 2745 & 307.2& 372.6\\ \hline
Code lines (test)    & 95.3\% & 1 & 1969 & 59.7& 152.2\\ \hline
Code tokens (test)   & 99.8\% & 1 & 917  & 15.8& 10.3\\
\hline
\end{tabular}%
}
\end{table}

\subsection{Experimental Settings}

\textbf{Dataset Partitioning.} 
To ensure a fair and reliable evaluation, we partitioned the dataset into training, validation, and test sets using an 8:1:1 ratio with stratified sampling. This approach preserves the original distribution of vulnerability types (i.e., CWE-IDs) across all subsets, thereby maintaining consistency in class proportions relative to the full dataset. Preserving this distribution is crucial for effective model training and generalization, particularly in the presence of class imbalance. Such stratified partitioning is a widely accepted practice in vulnerability analysis research~\cite{herbold2017jit,cito2023expert}.

\textbf{Hyperparameter Configuration.} 
Following established best practices~\cite{feng2020codebert,guo2020graphcodebert,wang2021codet5}, we employed the AdamW optimizer and adopted a cosine learning rate decay schedule over 100 training epochs. The batch size was set to 32. To identify optimal training configurations, we conducted an automated grid search based on validation performance, and this search process yielded the following hyperparameter settings:
\begin{itemize}
    \item Learning rate: $5 \times 10^{-6}$,
    \item PGD perturbation magnitude $\epsilon$: 0.02,
    \item Dropout rate: 0.2,
    \item Gradient clipping norm: 1.0.
\end{itemize}


\textbf{Running Environment.} Model training and evaluation were conducted on a high-performance workstation equipped with a 13th Gen Intel Core i5-13600K processor (14-core, 20-thread architecture), 64 GB of RAM, and an NVIDIA GeForce RTX 4090 GPU with 24 GB of memory. 

\section{Result Analysis}
\label{sec:Result}
\subsection{RQ1: Comparison with Baselines}
\subsubsection{Vulnerability Type Prediction}

\textbf{Approach.} 
To address this RQ, we evaluate the effectiveness of our proposed approach on the VTP task. We adopt four widely used evaluation metrics: Accuracy (Acc), Precision (Prec), Recall (Rec), and F1-score (F1), with F1-score serving as the primary metric due to its robustness in the presence of the class imbalance issue~\cite{li2023survey, zhou2023comprehensive}.

We compare EDAT-MTL with two state-of-the-art baselines: VulExplainer~\cite{fu2023vulexplainer} and LIVABLE~\cite{wen2024livable}. All models are evaluated under the same experimental settings on the Big-Vul dataset~\cite{fan2020ac}, including identical data splits and hyperparameter settings to ensure a fair comparison. For reproducibility, we utilize the implementations shared by the original studies.

\textbf{Result.} Table~\ref{tab:RQ1-1} presents the comparison results for vulnerability type prediction across three backbone code models: CodeBERT~\cite{feng2020codebert}, GraphCodeBERT~\cite{guo2020graphcodebert}, and CodeT5~\cite{wang2021codet5}. Based on the results in the table, we observe that our proposed approach consistently outperforms all baselines across all evaluation metrics.
Specifically, with the CodeT5 backbone, EDAT-MLT achieves the highest F1-score of 0.7476, surpassing the strongest baseline, LIVABLE, by 0.2007. On the GraphCodeBERT backbone, EDAT-MLT achieves an F1-score of 0.7394, outperforming VulExplainer by 0.2375. Using CodeBERT, it reaches an F1-score of 0.7281, exceeding VulExplainer by 0.2186.
These performance gains can be attributed to two key factors. First, EDAT generates adversarial variants at the embedding level, guided by identifier semantics and constrained by AST/PDG structures, which enables the model to learn vulnerability-relevant patterns beyond shallow lexical features. Second, the encoder-decoder architecture of CodeT5, coupled with its pretraining on a more diverse and comprehensive corpus, enhances its generalization ability to rare and underrepresented CWE types.

\begin{table}[H]
\centering
\scriptsize
\caption{Comparison results with baselines on the vulnerability type prediction task}
\label{tab:RQ1-1}
\setlength{\tabcolsep}{2pt} 

\resizebox{\linewidth}{!}{
\begin{tabular}{@{}l c c c c@{}}
\hline
\textbf{Model Configuration} & \textbf{Acc} & \textbf{Prec} & \textbf{Rec} & \textbf{F1} \\
\hline
VulExplainer-CodeBERT & 0.5211 & 0.5026 & 0.5234 & 0.5095 \\
VulExplainer-GraphCodeBERT & 0.5133 & 0.4951 & 0.5155 & 0.5019 \\
VulExplainer-CodeT5 & 0.5107 & 0.4925 & 0.5129 & 0.4993 \\
Livable & 0.5453 & 0.5478 & 0.5681 & 0.5469 \\
\cdashline{1-5}[1pt/1pt]
CodeBERT (Our) & 0.7318 & 0.7155 & 0.7329 & 0.7281 \\
GraphCodeBERT (Our) & 0.7462 & 0.7217 & 0.7409 & 0.7394 \\
CodeT5 (Our) & 0.7533 & 0.7248 & 0.7612 & 0.7476 \\
\hline
\end{tabular}
}
\end{table}

\begin{tcolorbox}[colback=gray!10, colframe=black, boxrule=0.8pt]
\textbf{Answer to RQ1-1:} Our approach outperforms the baselines in vulnerability type prediction, achieving up to 24.83 percentage points of improvement. This demonstrates the strong generalization capability of our proposed approach, particularly in enhancing recall for long-tail vulnerability types, which is critical for real-world security assurance.
\end{tcolorbox}

\subsubsection{Line-level Vulnerability Detection}

\textbf{Approach.} 
To answer this RQ, we evaluate the effectiveness of our proposed approach on the task of line-level vulnerability detection. We adopt five widely used evaluation metrics: Top-5 Accuracy (Top-5 ACC), Top-10 Accuracy (Top-10 ACC), Recall at Top 20\% Lines of Code (R@20\% LOC), Effort to reach 20\% Recall (E@20\% R), and Initial False Alarm (IFA). These metrics are chosen to capture both the precision of vulnerability localization and the manual inspection effort required—two critical aspects for assessing the practical utility of line-level vulnerability detection in real-world software engineering scenarios.

We compare EDAT-MTL with two state-of-the-art baseline methods, LineVD~\cite{hin2022linevd} and LineVul~\cite{fu2022linevul}, to evaluate its performance on the line-level vulnerability detection task. All approaches are assessed on the Big-Vul dataset~\cite{fan2020ac} using identical data splits and consistent training configurations to ensure a fair comparison. The evaluation is conducted across three backbone code models: CodeBERT~\cite{feng2020codebert}, GraphCodeBERT~\cite{guo2020graphcodebert}, and CodeT5~\cite{wang2021codet5}. To guarantee reproducibility and a fair comparison, we adopt the implementations shared by the original studies.

\textbf{Result.} Table~\ref{tab:RQ1-2} presents the comparison results. Based on these results, we find that using CodeBERT as the backbone model, EDAT-MLT outperforms LineVD and LineVul across all metrics, achieving 0.6047 in Top-5 ACC, 0.7328 in Top-10 ACC, 0.6519 in R@20\% LOC, and 2.7913 in IFA. These results reflect an improvement over LineVD, whose Top-5 ACC is 0.2350 and IFA is 12.4591, and LineVul, whose Top-5 ACC is 0.3862 and IFA is 11.5780.

Similar performance improvements can be observed on GraphCodeBERT and CodeT5. On GraphCodeBERT, EDAT-MLT achieves the highest R@20\% LOC of 0.7031 and the lowest IFA of 2.2546, benefiting from the encoder’s ability to model variable dependencies through program graphs. On CodeT5, EDAT-MLT obtains a Top-10 ACC of 0.8109 and competitive performance across all other metrics, demonstrating generalizability. These improvements can be attributed to two factors: (1) the EDAT module introduces adversarial perturbations constrained by AST/PDG structures, enabling the model to detect subtle code semantics related to vulnerability patterns; and (2) the MTL module leverages type identification guidance to improve line-level detection particularly where only a small portion of code lines are labeled.

\begin{table}[H]
\centering
\scriptsize
\caption{Comparison results with baselines on the line-level vulnerability detection task}
\label{tab:RQ1-2}
\setlength{\tabcolsep}{3pt}

\resizebox{\linewidth}{!}{
\begin{tabular}{@{}l c c c c c@{}}
\hline
\textbf{Configuration} & \textbf{Top-5 ACC} & \textbf{Top-10 ACC} & \textbf{R@20\% LOC}& \textbf{E@20\% R} & \textbf{IFA} \\
\hline
LineVD-CodeBERT & 0.2350 & 0.3751 & 0.3392 & 0.1768 & 12.4591\\
LineVul-CodeBERT & 0.3862 & 0.4864 & 0.3409 & 0.1968 & 11.5780\\
\cdashline{1-6}[1pt/1pt]
CodeBERT (Our) & 0.6047 & 0.7328 & 0.6519 & 0.0097 & 2.7913\\
GraphCodeBERT (Our) & 0.6356 & 0.7742 & 0.7031 & 0.0084 & 2.2546\\
CodeT5 (Our) & 0.6814 & 0.8109 & 0.7385 & 0.0078 & 1.9871\\
\hline
\end{tabular}
}
\end{table}

\begin{tcolorbox}[colback=gray!10, colframe=black, boxrule=0.8pt]
\textbf{Answer to RQ1-2:} Our approach consistently outperforms the baselines in line-level vulnerability detection, yielding substantial improvements in in Top-5/10 ACC, R@20\% LOC, and IFA. These results demonstrate the effectiveness of EDAT-MTL in accurately localizing vulnerabilities while reducing false positives, thereby enhancing its practical applicability.
\end{tcolorbox}


\subsection{RQ2: Influence of EDAT Module}
\subsubsection{Vulnerability Type Prediction}

\textbf{Approach.}
To answer this RQ, we conduct an ablation study to evaluate the effectiveness of our designed EDAT module on the task of vulnerability type classification. The goal is to determine whether EDAT consistently improves model performance across different code pre-trained models. Specifically, we compare the performance of three representative code models under two settings: with and without EDAT (w/o EDAT).
For each model, EDAT is implemented using Projected Gradient Descent, with a perturbation budget of $\epsilon$ = 0.03 applied to identifier embeddings over $n\_steps$ = 3. Progressive perturbation scheduling is used across training epochs. Evaluation is conducted on the Big-Vul dataset, using Accuracy, Precision, Recall, and F1-score as metrics, where F1-score serves as the core metric due to its robustness for the dataset with the class imbalance issue.



\textbf{Result.}
As shown in Table~\ref{tab:RQ2-1}, incorporating the EDAT module consistently enhances performance across all three pre-trained code models. On CodeBERT, EDAT boosts Accuracy from 0.6021 to 0.7318 and F1-score from 0.5965 to 0.7281, achieving an absolute F1-score improvement of 0.1316. For GraphCodeBERT, EDAT raises the F1-score from 0.6258 to 0.7394. The most significant improvement is observed on CodeT5, where the F1-score increases from 0.6427 to 0.7476.

These performance gains can be attributed to two primary factors. First, EDAT introduces gradient-guided perturbations to identifier tokens under AST/PDG constraints, ensuring syntactic correctness while promoting semantic diversity. This encourages models to attend to vulnerability-relevant logic rather than relying on superficial lexical cues. Second, training with adversarial examples compels models to learn more robust semantic representations, particularly for rare or underrepresented CWE types. As a result, the models exhibit stronger generalization and greater resilience to code obfuscation. Moreover, the effectiveness of EDAT is partially architecture-dependent. While all three models benefit from adversarial training, CodeT5 achieves the highest performance due to its encoder-decoder architecture and broader pre-training corpus, which better equips it to capture long-range semantic dependencies critical for both understanding and generation tasks.

\begin{table}[H]
\centering
\scriptsize  
\caption{Performance impact of the EDAT module on vulnerability type prediction}
\label{tab:RQ2-1}
\setlength{\tabcolsep}{3pt} 

\resizebox{\linewidth}{!}{ 
\begin{tabular}{@{}l c c c c@{}}
\hline
\textbf{Model Configuration} & \textbf{Acc} & \textbf{Prec} & \textbf{Rec} & \textbf{F1} \\
\hline
w/o EDAT (CodeBERT) & 0.6021 & 0.5888 & 0.6143 & 0.5965 \\
EDAT (CodeBERT) & 0.7318 & 0.7155 & 0.7329 & 0.7281 \\
\cdashline{1-5}[1.5pt/1.5pt]  
w/o EDAT (GraphCodeBERT) & 0.6416 & 0.6071 & 0.6539 & 0.6258 \\
EDAT (GraphCodeBERT) & 0.7462 & 0.7217 & 0.7409 & 0.7394 \\
\cdashline{1-5}[1.5pt/1.5pt]  
w/o EDAT (CodeT5) & 0.6592 & 0.6354 & 0.6683 & 0.6427 \\
EDAT (CodeT5) & 0.7533 & 0.7248 & 0.7612 & 0.7476 \\
\hline
\end{tabular}
}
\end{table}

\begin{tcolorbox}[colback=gray!10, colframe=black, boxrule=0.8pt]
\textbf{Answer to RQ2-1:} 
The ablation results demonstrate that EDAT consistently improves model performance across CodeBERT, GraphCodeBERT, and CodeT5, with the highest F1-score of 0.7476 achieved on CodeT5. These improvements validate EDAT’s effectiveness in generating syntactically valid and semantically meaningful adversarial variants, thereby enhancing both recall and overall robustness. 
\end{tcolorbox}

\subsubsection{Line-level Vulnerability Detection}

\textbf{Approach.} To assess the impact of the proposed EDAT module on line-level vulnerability detection, we conduct an ablation study using the Big-Vul dataset~\cite{fan2020ac} and three code pre-trained models: CodeBERT~\cite{feng2020codebert}, GraphCodeBERT~\cite{guo2020graphcodebert}, and CodeT5~\cite{wang2021codet5}. For each model, we also compare two variants: with and without EDAT (w/o EDAT).
For the EDAT module, adversarial perturbations are applied to identifier embeddings using the Projected Gradient Descent (PGD) method, with a perturbation budget of $\epsilon = 0.03$ and $n_steps = 3$. These perturbations are scheduled progressively over the first two training epochs.

To ensure a fair comparison, all variants are trained under the same hyperparameter settings and data splits. We adopt five widely used line-level detection metrics to evaluate both accuracy and inspection effort: Top-5 ACC, Top-10 ACC, R@20\% LOC, E@20\% R, and IFA. These metrics are selected for their ability to jointly reflect the precision and practicality of line-level detection models in real-world scenarios.

\textbf{Result.} 
Table~\ref{tab:RQ2-2} presents the performance comparison results. On CodeBERT, EDAT consistently improves Top-10 ACC from 0.6213 to 0.7328 and reduces IFA from 6.2145 to 2.7913. Notably, R@20\% LOC increases from 0.5582 to 0.6519, indicating a significant reduction in manual inspection effort required to identify vulnerable lines.
GraphCodeBERT benefits even more from EDAT, with Top-10 ACC increasing from 0.6522 to 0.7742 and R@20\% LOC from 0.5906 to 0.7031. These results demonstrate EDAT’s strong compatibility with graph-based architectures, which model data-flow relationships and help the model focus on semantically important structures rather than superficial lexical patterns.
Although the relative improvements on CodeT5 are smaller, the model still achieves strong results with EDAT: Top-10 ACC reaches 0.8109, and IFA decreases to 1.9871. This shows that EDAT can still yield meaningful gains even on models structurally optimized for generation tasks.
These improvements can be attributed to two main factors: (1) EDAT applies perturbations selectively to identifier embeddings and constrains them via AST/PDG structures, ensuring semantic diversity while maintaining syntactic validity; and (2) adversarial training promotes deeper semantic learning, enhancing the model's ability to generalize to rare or atypical vulnerability patterns.

\begin{table}[H]
\centering
\scriptsize
\caption{Performance impact of the EDAT module on  line-level vulnerability detection}
\label{tab:RQ2-2}
\setlength{\tabcolsep}{3pt}
\resizebox{\linewidth}{!}{
\begin{tabular}{@{}l c c c c c c@{}}
\hline
\textbf{Configuration} & \textbf{Top-5 ACC}$\uparrow$   & \textbf{Top-10 ACC}$\uparrow$ & \textbf{R@20\% LOC}$\uparrow$ & \textbf{E@20\% R}$\downarrow$ & \textbf{IFA}$\downarrow$ \\
\hline
w/o EDAT (CodeBERT) & 0.5174 & 0.6213 & 0.5582 & 0.0112 & 6.2145 \\
EDAT (CodeBERT) & 0.6047 & 0.7328 & 0.6519 & 0.0097 & 2.7913 \\
\cdashline{1-6}[1.5pt/1.5pt]
w/o EDAT (GraphCodeBERT) & 0.5491 & 0.6522 & 0.5906 & 0.0097 & 5.5964 \\
EDAT (GraphCodeBERT) & 0.6356 & 0.7742 & 0.7031 & 0.0084 & 2.2546 \\
\cdashline{1-6}[1.5pt/1.5pt]
w/o EDAT (CodeT5) & 0.5867 & 0.6935 & 0.6278 & 0.0090 & 5.2896 \\
EDAT (CodeT5) & 0.6814 & 0.8109 & 0.7385 & 0.0078 & 1.9871 \\
\hline
\end{tabular}
}
\end{table}

\begin{tcolorbox}[colback=gray!10, colframe=black, boxrule=0.8pt]
\textbf{Answer to RQ2-2:} Using the EDAT module enhances line-level vulnerability detection across different code models by introducing perturbations specifically to identifier embeddings, guided by structural constraints. These targeted, embedding-level modifications encourage the model to focus on semantically meaningful features while reducing manual inspection effort. Among the evaluated models, CodeT5 benefits the most from this approach.
\end{tcolorbox}

\subsection{RQ3: Influence of MTL Module}

\textbf{Approach.} To evaluate the impact of multi-task learning on both vulnerability type prediction and line-level detection, we conduct ablation studies comparing our multi-task model with its single-task counterparts. Specifically, we assess these two variants across three pre-trained code models: CodeBERT, GraphCodeBERT, and CodeT5.

\begin{itemize}
  \item \textbf{Single-Task:} This variant is trained exclusively on the vulnerability type prediction task, without incorporating the line-level detection objective. It serves as a baseline to isolate the impact of joint multi-task optimization.
  \item \textbf{Multi-Task (Ours):} The model is trained to jointly learn both VTP and LVD tasks using a unified architecture and shared parameters.
\end{itemize}

All models are trained on the Big-Vul datasets~\cite{fan2020ac} using the same hyperparameter settings to ensure a fair comparison. Evaluation metrics for vulnerability type prediction include Accuracy, Precision, Recall, and F1-score, while line-level detection metrics include Top-5 Accuracy, Top-10 Accuracy, Recall at Top 20\% LOC, Effort to reach 20\% Recall, and Initial False Alarm. These metrics provide a comprehensive assessment of both detection effectiveness and inspection efficiency.

\textbf{Result.} 
Table~\ref{tab:RQ3} presents the comparative results between the multi-task and single-task variants. As observed, incorporating the multi-task learning (MTL) module consistently improves performance across all code models and both tasks. For example, on CodeBERT, MTL boosts the classification F1-score from 0.4857 to 0.5965, increases the Top-10 Accuracy from 0.3661 to 0.6213, and reduces IFA from 13.0261 to 6.2145. These improvements demonstrate that MTL enhances the model’s ability to learn richer semantic representations and effectively mitigates overfitting, particularly in scenarios with limited training data.

GraphCodeBERT also benefits from the MTL setting, with the F1-score increasing from 0.5185 to 0.6258, R@20\% LOC improving from 0.3527 to 0.5906, and IFA decreasing from 12.2400 to 5.5964. These results suggest that MTL helps the model better exploit the structural dependencies captured by graph-based code encoders, thereby improving generalization.

Among all the models, CodeT5 achieves the best overall performance, with an F1-score of 0.6427 and a Top-10 Accuracy of 0.6935. Its encoder-decoder architecture and broader pre-training corpus allow it to integrate auxiliary signals more flexibly, which proves particularly advantageous for line-level vulnerability detection. These findings confirm that MTL effectively captures complementary signals across tasks and architectures, leading to more robust and transferable representations.

\begin{table*}[htbp]
\centering
\caption{Comparison of multi-task and single-task models on vulnerability type prediction and line-level vulnerability detection}
\label{tab:RQ3}
\small
\setlength{\tabcolsep}{4pt} 
\renewcommand{\arraystretch}{1.15} 
\resizebox{\linewidth}{!}{
\begin{tabular}{@{}lccccccccc@{}}
\toprule
\textbf{Model} & \multicolumn{4}{c}{\textbf{Vulnerability Type Prediction}} & \multicolumn{5}{c}{\textbf{Line-level Vulnerability Detection}} \\
\cmidrule(lr){2-5} \cmidrule(lr){6-10}
 & Acc & Prec & Rec & F1 & Top-5 ACC$\uparrow$ & Top-10 ACC$\uparrow$ & R@20\%LOC$\uparrow$ & E@20\%R$\downarrow$ & IFA$\downarrow$ \\
\midrule
Single-CodeBERT & 0.4982 & 0.4764 & 0.5013 & 0.4857 & 0.2879 & 0.3661 & 0.3294 & 0.1392 & 13.0261 \\
Mult-CodeBERT & 0.6021 & 0.5888 & 0.6143 & 0.5965 & 0.5174 & 0.6213 & 0.5582 & 0.0112 & 6.2145 \\

\cdashline{1-10}[1.5pt/1.5pt]
Single-GraphCodeBERT & 0.5337 & 0.5099 & 0.5424 & 0.5185 & 0.3084 & 0.4958 & 0.3527 & 0.1212 & 12.2400 \\
Mult-GraphCodeBERT & 0.6416 & 0.6071 & 0.6539 & 0.6258  & 0.5491 & 0.6522 & 0.5906 & 0.0097 & 5.5964 \\

\cdashline{1-10}[1.5pt/1.5pt]
Single-CodeT5 & 0.5368 & 0.5342 & 0.5475 & 0.5456 & 0.3372 & 0.4203 & 0.3616 & 0.1128 & 11.8560 \\
Mult-CodeT5 & 0.6592 & 0.6354 & 0.6683 & 0.6427 & 0.5867 & 0.6935 & 0.6278 & 0.0090 & 5.2896 \\
\bottomrule
\end{tabular}
}
\end{table*}

\begin{tcolorbox}[colback=gray!10, colframe=black, boxrule=0.8pt]
\textbf{Answer to RQ3:} Multi-task learning consistently improves model performance on both vulnerability type prediction and line-level vulnerability detection. By sharing parameters and enabling joint optimization, MTL helps enhance the performance of these two correlated tasks simultaneously. Among the evaluated models, CodeT5 exhibits the best performance improvements, which can be attributed to its larger pre-training corpus and more advanced model architecture.
\end{tcolorbox}

\section{Discussions}
\label{sec:Discussions}

\subsection{Influence of Hyperparameter Setting}

To investigate the impact of key hyperparameters on model performance, we conducted a systematic sensitivity analysis using grid search. This analysis examines how variations in learning rate, dropout rate, and PGD perturbation magnitude ($\epsilon$) affect the performance of our approach on both the VTP and LVD tasks. The detailed results are presented in Table~\ref{tab:hyperparameter_codeT5}.

The analysis reveals that the learning rate plays a critical role in balancing convergence and generalization. A learning rate that is too high (e.g., 1e-5) leads to training instability (although recall remains comparable, 
line-level detection precision is notably degraded, as reflected by an increase in IFA from 1.9871 to 2.2450). On the other hand, an overly conservative rate (e.g., 1e-6) hinders convergence and risks the model being trapped in sub-optimal solutions. Similarly, the choice of PGD perturbation magnitude ($\epsilon$) involves a delicate trade-off: a small value (e.g., 0.01) fails to induce meaningful robustness, while a large value (e.g., 0.05) introduces excessive semantic noise, resulting in performance degradation across all metrics.

The final parameter configuration (i.e., a learning rate of 5e-6, a dropout rate of 0.2, and an $\epsilon$ value of 0.02) was selected not to optimize a single metric, but to achieve the best overall balance across multiple tasks and evaluation indicators. This setup yields the highest F1-score (74.76\%) and Recall (76.12\%) on the VTP task, while also attaining the best Top-10 Accuracy (0.8109) and the lowest IFA (1.9871) on the LVD task. These results indicate an optimal trade-off between model generalization and detection performance.


\begin{table}[H]
\centering
\scriptsize
\caption{Hyperparameter Sensitivity Analysis}
\label{tab:hyperparameter_codeT5}
\setlength{\tabcolsep}{4pt} 

\begin{tabular}{@{}llcccc@{}}
\toprule
\multirow{2}{*}{\textbf{Hyperparameter}} & \multirow{2}{*}{\textbf{Value}} & \multicolumn{2}{c}{\textbf{VTP Performance}} & \multicolumn{2}{c}{\textbf{LVD Performance}} \\
\cmidrule(lr){3-4} \cmidrule(lr){5-6}
& & \textbf{F1-Score(\%)}& \textbf{Recall(\%)}& \textbf{Top-10 Acc.} & \textbf{IFA↓} \\
\midrule
\multirow{3}{*}{Learning Rate} & 1e-6                   & 72.45& 70.13& 0.7941& 2.1651\\
                               & \textbf{5e-6 (Optimal)}& \textbf{74.76}& \textbf{76.12}& \textbf{0.8109}& \textbf{1.9871}\\
                               & 1e-5                   & 71.56& 73.24& 0.7867& 2.2450\\
\midrule
\multirow{3}{*}{Dropout Rate}  & 0.0                    & 73.55& 72.67& 0.8013& 2.1503\\
                               & \textbf{0.2 (Optimal)} & \textbf{74.76}& \textbf{76.12}& \textbf{0.8109}& \textbf{1.9871}\\
                               & 0.4                    & 72.14& 71.89& 0.7945& 2.3110\\
\midrule
\multirow{3}{*}{PGD $\epsilon$}& 0.01                   & 73.42& 72.65& 0.8056& 2.2104\\
                               & \textbf{0.02 (Optimal)}& \textbf{74.76}& \textbf{76.12}& \textbf{0.8109}& \textbf{1.9871}\\
                               & 0.05                   & 71.14& 69.67& 0.7883& 2.2460\\
\bottomrule
\multicolumn{6}{l}{\textit{Results are reported on the CodeT5 backbone.}}
\end{tabular}
\end{table}

\subsection{Threats to Validity}
\label{Threats to Validity}

In this subsection, we analyze potential threats to the validity of our study.
For each threat, we show the corresponding mitigation methods.

\textbf{Internal Validity.}
This threat concerns potential errors that may arise during the reproduction of the selected baselines and our proposed approach. To mitigate this risk, we used the official implementations provided by the original studies rather than reimplementing the models ourselves, thereby minimizing the chance of implementation mistakes. Furthermore, we adopted the default hyperparameters and experimental protocols recommended in the original papers to ensure a fair and consistent comparison. To ensure the correctness of our proposed approach, we conducted careful code inspections throughout the implementation process.

\textbf{External Validity.}  
This threat concerns the generalizability of our results. To validate the effectiveness of our approach, we used the Big-Vul dataset, which is widely adopted in existing vulnerability detection research~\cite{lu2023assessing,ren2025improving,liu2024making,ren2024prorlearn,wang2025sift,lu2024grace} and provides the necessary annotations for both tasks. However, our experiments are conducted primarily on C++ code, and the findings may not generalize well to other programming languages with different paradigms, syntax, and ecosystem characteristics, such as Python or Java.

\textbf{Construct Validity.}
The threat to construct validity concerns the performance metrics we employed. To mitigate this threat, we adopted widely used evaluation metrics for both the VTP and LVD tasks. For the VTP task, the selected metrics primarily assess overall prediction accuracy and the trade-off between correctly identifying vulnerable instances and minimizing false positives. For the LVD task, the evaluation focuses on the model’s ability to accurately rank and localize vulnerable lines within code, as well as its efficiency in achieving high recall with minimal inspection effort.


\section{Related Work}
\label{sec:Related}

Our work targets two core tasks in software vulnerability detection and understanding: vulnerability type prediction and line-level vulnerability detection. In this section, we first review prior work related to these two tasks, and then highlight the novel contributions of our study.

\textbf{Vulnerability Type Prediction.}  
Vulnerability type prediction is the task of identifying security flaws in source code and categorizing them into standardized taxonomies, such as CWE-IDs. Traditional methods rely on handcrafted features or shallow text representations (e.g., Bag-of-Words or TF-IDF)~\cite{chakraborty2021deep,bilgin2020vulnerability}, which lack the capacity to capture deep syntactic and semantic structures in code~\cite{szabo2023new}. Recent advances in deep learning have introduced code representation learning techniques based on abstract syntax trees~\cite{sun2023abstract}, control flow graphs~\cite{conrado2023bounded}, and graph-based encoders~\cite{qiu2024vulnerability}, which enable hierarchical modeling of vulnerability-indicative patterns. These approaches have demonstrated notable success in improving the accuracy of CWE classification~\cite{chakraborty2021deep}.

\textbf{Line-level Vulnerability Detection.}  
Line-level vulnerability detection focuses on identifying the exact code lines or functions responsible for Vulnerabilities. Traditional static analyzers and rule-based tools (such as FlawFinder and RATS) tend to exhibit high false positive rates due to reliance on pattern matching~\cite{fu2022linevul}. Machine learning models using token-level features also struggle with semantic understanding. To address these issues, recent studies like LineVul~\cite{fu2022linevul} and IVDetect~\cite{li2021vulnerability} adopt Transformer architectures enhanced with graph neural networks to better capture data/control-flow semantics. These models improve the accuracy of line-level vulnerability detection by effectively leveraging structural information embedded in the code.

\textbf{Novelty of Our Study.}
Despite recent progress in vulnerability detection and understanding, existing studies face two major limitations: (1) the scarcity of high-quality labeled data, which hampers the training of robust models, and (2) the neglect of task correlations between different but related objectives such as VTP and LVD, which are often treated in isolation. To address these challenges, we propose a unified framework that integrates two novel components. First, the Embedding-layer Driven Adversarial Training module introduces semantically meaningful perturbations to identifier embeddings under syntactic constraints, enhancing model robustness and generalization. Second, the Multi-Task Learning module jointly learns VTP and LVD in a shared architecture, allowing knowledge transfer across tasks and promoting richer semantic representations. Experimental results on the widely-used Big-Vul dataset demonstrate that our approach can achieve an obvious performance improvement over state-of-the-art baselines on both tasks. Furthermore, ablation studies validate the individual effectiveness of the EDAT and MTL modules.

\section{Conclusion and Future Work}
\label{sec:Conclusion}

In this study, we investigate the challenges of software vulnerability detection and understanding by focusing on two complementary tasks: vulnerability type prediction and line-level vulnerability detection. While prior work has explored these tasks separately, our study highlights the benefit of a unified treatment that leverages both task synergy and robust training strategies. We introduce two key components: an embedding-layer adversarial training module and a multi-task learning module, designed to address the limitations of data scarcity and isolated task learning. Through comprehensive experiments on the Big-Vul dataset, we demonstrate that our approach consistently improves model performance across multiple code pre-trained models, outperforming state-of-the-art baselines in both tasks. Additionally, ablation results confirm that both EDAT and MTL individually contribute to the overall effectiveness of our proposed approach, particularly in enhancing generalization and reducing false positives. Our study provides an effective pathway to enhance both vulnerability type classification and line-level detection accuracy, marking an important step forward and a fruitful area for continued exploration in automated vulnerability analysis.

In the future, we plan to extend our proposed approach to other programming languages by collecting relevant datasets for model training. Additionally, we aim to explore more effective data augmentation techniques and advanced multi-task learning strategies to further enhance the performance of our approach.

\section*{Acknowledgement}

Siyu Chen and Jiongyi Yang have contributed equally to this work, and they are co-first authors. Xiang Chen is the corresponding author.
This research was partially supported by the National Natural Science Foundation of China (Grant No. 61202006), the Open Project of State Key Laboratory for Novel Software Technology at Nanjing University under (Grant No. KFKT2024B21), and Postgraduate Research \& Practice Innovation Program of Jiangsu Province (Grant No. SJCX25\_2003).

\section*{Declaration of Competing Interests}
The authors declare that they have no known competing financial interests or personal relationships that could have appeared to influence the work reported in this paper.
	
\section*{CRediT Authorship Contribution Statement}

\textbf{Siyu Chen}: Conceptualization, Data curation, Methodology,
Software, Validation, Writing – review \& editing.
\textbf{Jiongyi Yang}: Conceptualization, Data curation, Methodology,
Software, Validation, Writing – review \& editing.
\textbf{Xiang Chen:} Conceptualization, Methodology, Writing -review \& editing, Supervision.
\textbf{Menglin Zheng:} Data curation, Software, Validation.
\textbf{Minnan Wei:} Data curation, Software, Validation.
\textbf{Xiaolin Ju:} Methodology, Writing -review \& editing.

\bibliography{mylib}
\bibliographystyle{elsarticle}

\vspace{1cm}

\noindent\textbf{Siyu Chen} is currently pursuing the Master degree at the School of Artificial Intelligence and Computer Science, Nan-tong University. Her research interests include software vulnerability analysis.

\par
\vspace{1cm}

\noindent\textbf{Jiongyi Yang} is currently pursuing the undergraduate degree at the School of Artificial Intelligence and Computer Science, Nantong University. His research interests focus on software vulnerability analysis.

\par
\vspace{1cm}
  
\noindent\textbf{Xiang Chen} received the B.Sc. degree in information management and systems from Xi'an Jiaotong University, China in 2002. Then he received his M.Sc., and Ph.D. degrees in computer software and theory from Nanjing University, China in 2008 and 2011 respectively. He is currently an Associate Professor at the School of Artificial Intelligence and Computer Science, Nantong University. He has authored or co-authored more than 160 papers in refereed journals or conferences, such as IEEE Transactions on Software Engineering, ACM Transactions on Software Engineering and Methodology, Empirical Software Engineering, Information and Software Technology, Journal of Systems and Software, Software Testing, Verification and Reliability, Journal of Software: Evolution and Process, International Conference on Software Engineering (ICSE), International Conference on the Foundations of Software Engineering (FSE), International Symposium on Software Testing and Analysis (ISSTA), International Conference Automated Software Engineering (ASE), International Conference on Software Maintenance and Evolution (ICSME), International Conference on Program Comprehension (ICPC), and International Conference on Software Analysis, Evolution and Reengineering (SANER). His research interests include software engineering, in particular large language models for software engineering, software testing and maintenance, software repository mining, and empirical software engineering. He received two ACM SIGSOFT distinguished paper awards in ICSE 2021 and ICPC 2023. He is the editorial board member of Information and Software Technology. More information can be found at: 

https://xchencs.github.io/index.html.

\par
\vspace{1cm}

\noindent\textbf{Menglin Zheng} is currently pursuing a Bachelor’s degree in Software Engineering at the School of Artificial Intelligence and Computer Science, Nantong University, with a focus on Artificial Intelligence and Multimodal Analysis.

\par
\vspace{1cm}

\noindent\textbf{Minnan Wei} is currently pursuing the Master degree at the School of Artificial Intelligence and Computer Science, Nantong University. His research interests include competitive program generation and vulnerability detection.

\par
\vspace{1cm}

\noindent\textbf{Xiaolin Ju} was born in April 1976. He received the B.S. degree in information science from Wuhan University in 1998, the M.Sc. degree in computer science from Southeast University in 2004, and the Ph.D. degree in computer science from the Chinese University of Mining Technology in 2014. He is currently an associate professor at the School of Information Science and Technology, Nantong University, Nantong, China. His current research interests include software testing, such as collective intelligence, deep learning testing and optimization, and software defects analysis.

\end{document}